\documentstyle[12pt]{article}
\input epsf
\begin{document}

\newcommand{\als}{\alpha_s(m_Z)}
\newcommand{\cacf}{\frac{C_A}{C_F}}
\newcommand{\pho}{\tilde{\gamma}}
\newcommand{\gl}{\tilde{g}}
\newcommand{\sneu}{\tilde{\nu}}
\newcommand{\st}{\tilde{t}}
\newcommand{\sq}{\tilde{q}}
\newcommand{\se}{\tilde{e}}
\newcommand{\ch}{\chi^{\pm}}
\newcommand{\neut}{\chi^{0}}
\newcommand{\gsi}{\,\raisebox{-0.13cm}{$\stackrel{\textstyle>}
{\textstyle\sim}$}\,}
\newcommand{\lsi}{\,\raisebox{-0.13cm}{$\stackrel{\textstyle<}
{\textstyle\sim}$}\,}
\newcommand{\be}{\begin{equation}} \newcommand{\ee}{\end {equation}}

\rightline{RU-97-22} 
\baselineskip=18pt \vskip 0.7in 
\begin{center} {\bf \LARGE Are Light Gluinos Dead? }\\ 
\vspace*{0.9in} 
{\large Glennys R. Farrar}\footnote{Invited talk at Rencontres de la
  Vallee d'Aoste, La Thuile, March 1997.  Research supported in part
  by NSF-PHY-94-23002.} \\ 
\vspace{.1in} 
{\it Department of Physics and Astronomy \\ Rutgers
University, Piscataway, NJ 08855, USA}\\ 
\end{center} 
\vspace*{0.2in}
\vskip 0.3in 

{\bf Abstract:} 
Not yet.  ALEPH's recent exclusion limit employs an aggressive
determination of theoretical uncertainties using a simplified
application of the Bayesian method.  The validity of their
analysis can be evaluated by its further implications, such
as contradicting the existence a b quark and requiring relations
between hadronic event-shape observables which are not
observed. Traditional error estimation methods result in a much larger
estimate for the theoretical uncertainties.  This puts the ALEPH and
also Csikor-Fodor limits at the $\sim 1~ \sigma$ level for the very
light gluino scenario.  A recent astrophysical result implies direct
searches will be more difficult than previously anticipated, adding to
the importance of reducing the QCD uncertainty in predictions
sensitive to indirect effects of light gluinos.  Some possible
indications in favor of a light gluino are noted.

\thispagestyle{empty} 
\newpage 
\addtocounter{page}{-1}
\newpage

All the LEP experiments have addressed the question of constraining
QCD color factors by high-statistics data on hadronic event shapes in
$Z^0$ decay.  See, e.g.,
\cite{opal:4j95,delphi:tagged4j,L3:ycut,aleph:lg} and references
therein. The agreement with QCD is entirely adequate, but  
the sensitivity to the effective number of flavors has not been
sufficient up to now to rule out the presence of a light gluino which
would increase the effective number of flavors from $\sim 5$ to $\sim
8$.  The strategy of the new ALEPH analysis\cite{aleph:lg} is to fit
{\it both} 4-jet angular distributions {\it and} the differential
2-jet rate\footnote{For each event, the value of the jet-definition
parameter $y_{ cut}$ at which the event changes from being a 2-jet
to a 3-jet event is determined.  This quantity is called $y_3$. The
differential 2-jet rate ($D_2$) is $\frac{1}{\sigma} \frac{d \sigma}{d
y_3}$.  We discuss ALEPH's fit B, which fixes $\frac{C_A}{C_F}$ to the
QCD value.} to simultaneously constrain $\alpha_s(m_Z)$ and $n_f$, the 
effective number of flavors.  Since every hadronic event contributes
to the differential two jet rate, it is well-determined statistically.
It is a first-order observable, i.e., its leading contribution is
$O(\alpha_s)$, so it is sensitive to the running of $\alpha_s$.  Thus
it feels $n_f $ through the coefficient of the $\beta$ function, $b_0
= \frac{11}{6}\frac{C_A}{C_F} - \frac{2}{3} n_f \frac{T_F}{C_F}$.  The
various 4-jet angular distributions are second-order variables, i.e.,
at the parton level their leading contribution is
$O(\alpha_s^2)$. Particular angular correlations enhance the
sensitivity to 4-fermion final states in comparison to the dominant $q
\bar{q} gg$ final states.  However 4-jet angular distributions are
less precisely determined statistically than is the differential 2-jet
rate.  The nice observation of ALEPH is that the shapes of the error
ellipses from the two types of measurements tend to be
complementary\footnote{For a more detailed discussion of this point
  see \cite{dissertori}.}, so combining the constraints could produce a
more precise determination than either one alone.  The final result
quoted by ALEPH from this analysis is\cite{aleph:lg}:  $\als = 0.1162
\pm 0.0012(stat) \pm 0.0040 (syst)$ and $n_f = 4.24 \pm 0.29(stat) \pm
1.15(syst)$.  The crucial question is whether their estimate of the
systematic uncertainty is realistic or overly optimistic.  

The fundamental issue is ALEPH's approach to assigning quantitative
systematic error bars.  For each source of systematic uncertainty
considered, several variations on the model are fit to the data,
resulting in a determination of $\als$ and $n_f$, with some $\chi^2$
for the fit.  They adopt what they characterize as a ``Bayesian point
of view'', which they describe as follows\cite{aleph:lg}:  ``The
Bayesian idea is that {\it a priori} all models can be considered 
equally well suited for usage in the analysis, but from a bad $\chi^2$
it is deduced that the {\it a posteriori} probability of such a model
is low, and therefore this model should get a small weight when
estimating the actual systematic error.  ...the systematic error
corresponds to the increase in $\chi^2$ by one.''  If all possible
models, or at least a complete unbiased sample of models is
explored by the analysis, the above ansatz should be valid on average
over many experimental fits.  The question which must be addressed is
whether the ansatz remains valid when restricted to the class of model
variations considered by ALEPH\footnote{For a discussion of Bayesian
methodology, see ref. \cite{dagostini:lec}.  Note that the formulae
used in ref. \cite{aleph:lg} to define $1 \sigma$ systematic errors
were developed by ALEPH and a derivation is not available in the
literature.  G. Dissertori, private communication.}. 

It is intuitively plausible that when data are being modeled from
first principles with a correct theory, and only the parameters of the
theory require specification, that the prescription outlined above
would give a correct estimation of errors.  Moving away from this
ideal but rarely realized case, one may have a situation in which the
{\it form} of some model function is known theoretically (e.g., the
power-dependence of pseudoscalar meson masses on the light-quark mass
in the $m_q \rightarrow 0$ limit of a lattice gauge theory
calculation) and only some parameters of the function are unknown.
Here it is also possible to rely on the Bayesian analysis as described
above, for aspects of the modeling which refer to the known
functional form.  

However in the situation most commonly encountered in practice,
the form of functions appearing in the model {\it are
not} known from first principles.  Examples are the functional forms
which are used to describe parton distribution functions or parton
hadronization.  When, as is usually the case, functional forms are
chosen for simplicity and convenience, and parameters tuned to fit a
diverse collection of data, naive application of the Bayesian approach
can lead to a completely wrong estimate of the actual theoretical
uncertainty.  A recent case in point is the production of high
$p_\perp$ jets observed by CDF which was anomalously large in
comparison to predictions of QCD with the then-standard gluon 
distribution functions\cite{cdf:hipjets}.  With the standard
functional form of the gluon distribution function, the QCD prediction
seemed very strongly constrained by other data.  However only a
mild generalization of the form was needed to fit all data including
the high-$p_\perp$ jets\cite{cteq4lq}.  

The most difficult situation of all is when one must model both the
conventional theory and the theory with some new degree of freedom.
In this case, to completely survey the model space as required for a
correct analysis, requires modeling and fitting {\it all} the physical
observables used to constrain the models, with and without the new
degree of freedom.  Here, since the form and parameters of 
hadronization models are tuned to agree with an enormous body of data,
the correct application of Bayesian principles minimally requires
generalizing Herwig and JETSET to include light gluinos and varying
all the parameters of the hadronization models to fit all data.  To
the extent that the underlying physics (e.g., non-perturbative QCD) is
well or poorly understood, the generalization of the modeling to new
degrees of freedom will or will not provide an adequate framework for
the Bayesian analysis.  

In addition to these non-perturbative issues, the perturbative QCD
predictions at the parton level depend on the renormalization scale
$\mu$, the matching scheme used to obtain resummed predictions, and
also on the parameter $y_{cut}$ and scheme used to define 4-jet
events.  The problem of fully exploring the model dependence
associated with these parameters is discussed below. 

Given that the full implementation of the Bayesian approach is
so demanding, we can ask whether it might still be valid in practice
when applied in the simplified way adopted by ALEPH.  Independent
experimental evidence will allow us to answer this question. 

Three sources of uncertainty dominate the ALEPH estimate of their
systematic error:  
\begin{itemize}
\item The uncertainty in the theoretical prediction for $D_2$, the
differential 2-jet rate, due to sensitivity to renormalization scale,
$\mu$. 
\item The uncertainty in the theoretical prediction for $D_2$ due to
  sensitivity to uncontrolled aspects of the resummation of large
  logarithms of $y_3$. 
\item The modeling of the hadronization of the partons.
\end{itemize} 
Two other uncertainties in the theoretical predictions which may be
important but which were not considered are:
\begin{itemize}
\item The uncertainty in the theoretical prediction for the
4-jet angular distributions due to sensitivity to renormalization
scale, $\mu$.
\item The uncertainty in the theoretical prediction for the
4-jet angular distribution due to large logarithms of $y_{cut}$, i.e.,
the dependence of the theoretical prediction on jet definition.
\end{itemize} 

\begin{figure}
\epsfxsize=\hsize
\epsffile{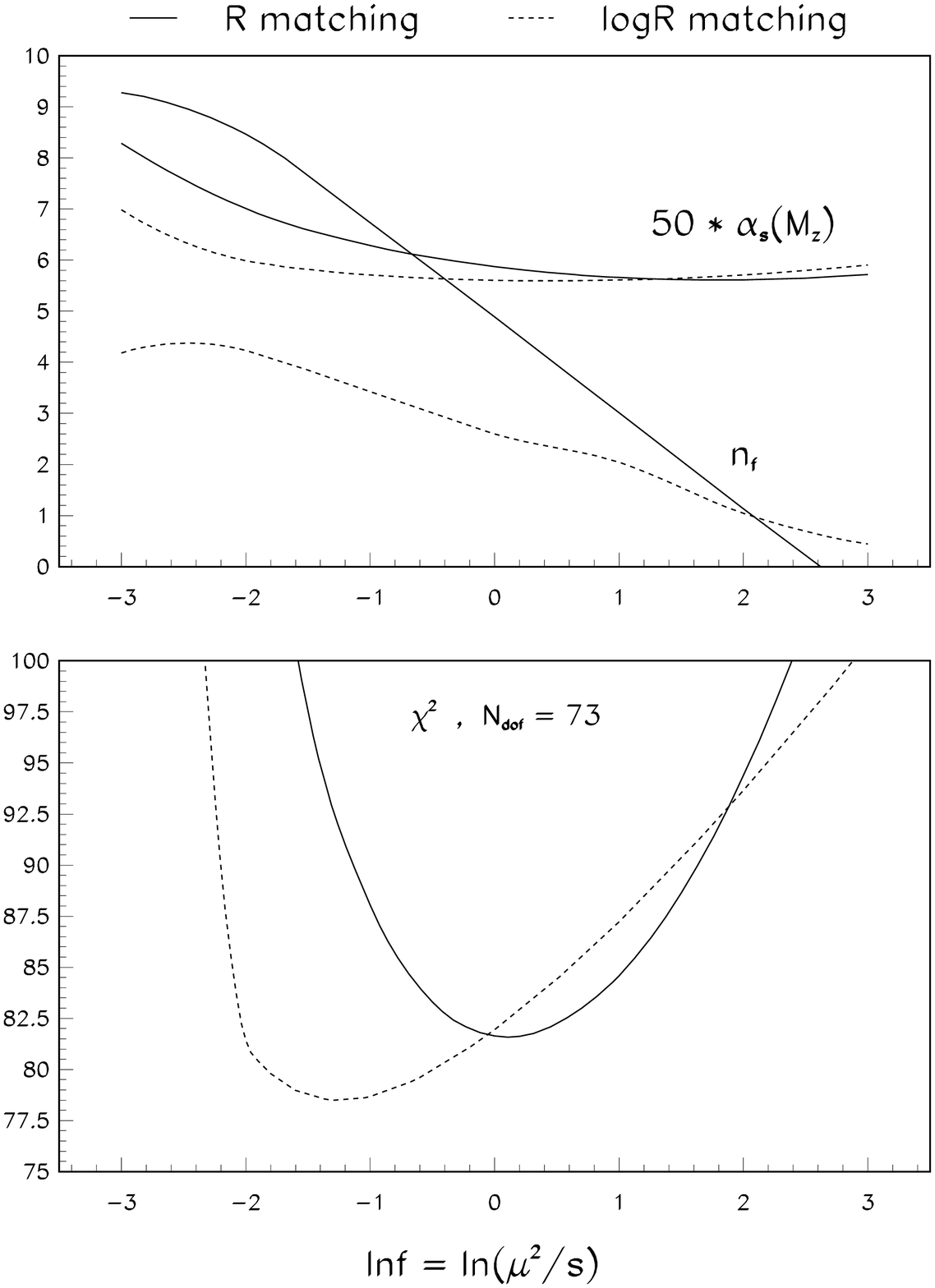}
\caption{Fit results with the R (solid) and logR (dashed) matching
  schemes as a function of $ln f \equiv ln \mu^2/s$, from \cite{aleph:lg}.}
\label{aleph:mudep}
\end{figure}

Two schemes for treating the effects of subleading logarithms of $y_3$
in the prediction for the differential 2-jet rate were considered,
R matching and log-R matching.  If subleading logs and higher order
corrections are unimportant, the schemes should give identical results
and should exhibit only weak scale dependence.  Fig. \ref{aleph:mudep}
from \cite{aleph:lg} shows their fit results for the two
schemes, along with the $\chi^2$ for each fit, as a function of $ln f
\equiv ln(\mu^2/m_Z^2)$.  It can be seen that the extracted value of
$n_f$ is sensitive to $\mu$, in both schemes.  Since 
$n_f$ is a physical quantity, it cannot depend on $\mu$ or scheme.

If one knew that the true prediction for the functional dependence on
$y_3$ of the differential 2-jet distribution were given by the
functional dependence predicted by one of the schemes, for some
particular value of $\mu$, then one could obtain the uncertainty
associated with the $\mu$ dependence by the Bayesian prescription, and
would get the correct mean value of the physical parameters by using
the scheme and $\mu$ value with the smallest $\chi^2$. ALEPH assumes
this is the case and takes the 1-sigma ``theoretical error'' (scale
uncertainty) to be$^{4}$ (c.f., formulae (15,16) of ref. \cite{aleph:lg}) 
\be
\Delta^{th} n_f = C ~{\rm max}_{[i]} \left( \frac{ | n_f^{[i]} - n_f^*
    |}{\sqrt{{\rm max} (1, | \chi^2_* - \chi^2_{[i]}|)}}\right)~~;~~ C
= {\rm max} \left( 1, \sqrt{\chi^2_*/N_{dof}} \right).
\label{aleph_sig}
\ee
$i$ labels the fit parameters of various models, with $*$ designating
the best fit.  Table \ref{aleph:tab3}, taken from Table 3 of ref. 
\cite{aleph:lg}, shows for the two schemes the best-fit and $\Delta
\chi^2 = \pm 1$ fits.  Rather than treating $i$ as a continuous variable
regarding the $\mu$ dependence shown in Fig. \ref{aleph:mudep}, they
only consider the discrete choices listed in Table \ref{aleph:tab3};
applying eqn (\ref{aleph_sig}), they find $\Delta^{th} =
1.01$.\footnote{Applying eqn (\ref{aleph_sig}) but considering $i$ to
  run over all $\mu$ gives $\Delta^{th} = 1.1$.}  

The best fit (log R matching with $ln f = -1.3$) gives $n_f =
3.68$.  However ALEPH averages the best-fit results of the log R and R
matching schemes and quotes $n_f = 4.24 \pm 1.01(th) \pm 0.45 (had)
\pm 0.27 (de) \pm 0.29 (stat)$. 

\begin{table}[htb]
\caption[]{Scheme and scale variations in ALEPH Fit B analysis,
  $N_{dof} = 73$. }  
\begin{tabular}{|l|r|r|r|c|}  \hline 

Variations & $\rm \alpha_s(M_Z)$ & $n_f$ & $\chi^2$  & $\mu/M_Z$ \\
  \hline
\multicolumn{5}{|c|}{Theoretical Prediction} \\ \hline

nominal: $\log$R, ln $f = -$1.3 \hspace*{.75in} & 0.1154 &
3.68 & 78.5 & 0.52\\  \hline

$\log$R, ln $f = -$1.8 & 0.1182 & 4.08 & 79.8 & 0.41\\ \hline

$\log$R, ln $f = -$0.6 & 0.1130 & 3.09 & 79.6 & 0.74\\ \hline
 
$R$, ln $f = -$0.5 & 0.1210 & 5.81 & 83.3 & 0.78\\ \hline 

$R$, ln $f$ = 0.0 & 0.1175 & 4.88 & 81.6 & 1.00\\ \hline

$R$, ln $f$ = 0.7 & 0.1141 & 3.57 & 83.0 & 1.42\\ \hline
\end{tabular}
\label{aleph:tab3}
\end{table}

The traditional prescription for estimating the $\pm 1 \sigma$
systematic error from scale uncertainty is to take it to be the range
of $n_f$ obtained by varying $\mu/\sqrt{s}$ over the range
[1/2,2]\footnote{In the recent analysis \cite{dixon_signer:R4}, the
  range [1/3,3] is also considered.}, i.e., $ln f$ between $\pm 1.4$.
From Fig. \ref {aleph:mudep} this gives $ n_f = 4.88 \pm 2.5(th)$, for
the R-matching scheme, and $n_f = 2.7 \pm 1(th)$ for the logR
scheme. The conservative conclusion is that the dependence of measured
distributions on $n_f$ is sensitive to truncation of the 
perturbation series (reflected in the sensitivity to $\mu$) and also
to subleading logarithms of $y_3$ (reflected in the sensitivity to
matching scheme) and therefore one cannot draw strong conclusions
until the theoretical predictions are improved.   

Fortunately, we have two pieces of independent empirical evidence
to test the validity of ALEPH's more aggressive estimate of the
scale and resummation scheme uncertainty. 

\begin{figure}
\epsfxsize=\hsize
\epsffile{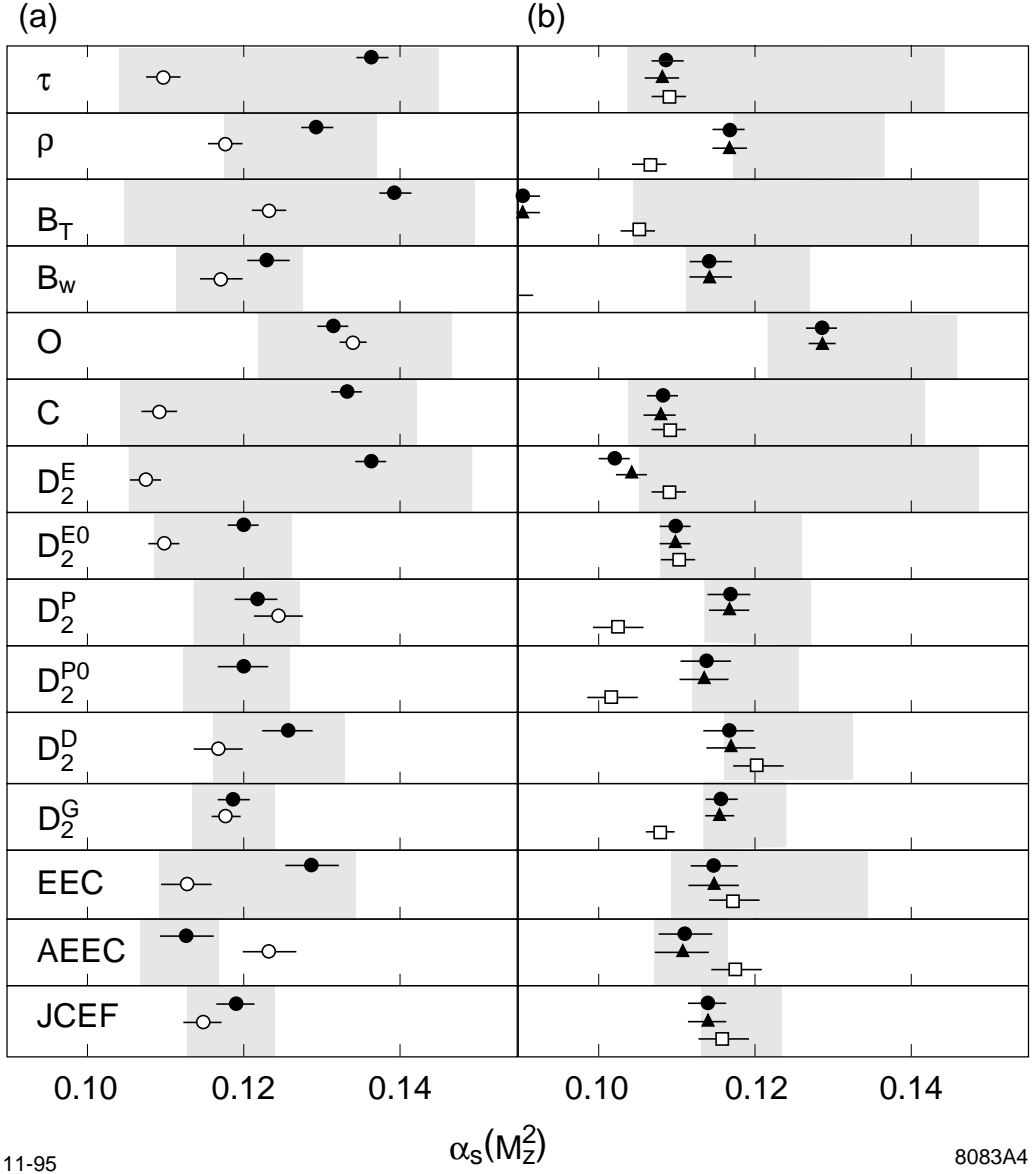}
\caption{(a) values of $\als$ from QCD fits to various event shape
  distributions using ``physical'' ($\mu = M_Z$) -- solid circles -- and
  experimentally-optimized scales -- open circles.  For each
  observable the shaded region indicates the total uncertainty
  estimated in Ref. \cite{sld:alphas}, dominated by the contribution
  from wide variation of the renormalization scale.  (b) shows similar
  comparisons for other scale-setting prescriptions:  Principle of
  Minimal Sensitivity (solid circles), Fastest Apparent Convergence
  (solid triangles) and Brodsky-Lepage-Mackenzie (open squares).
  Taken from ref. \cite{burrows:warsaw}.} 
\label{burrows1}
\end{figure}

First of all, consider the procedure for fixing the scale
  uncertainty.  As noted above, the underlying assumption is that
  there is some value of $\mu$ for which the net effect of neglected
  higher orders and subleading logs is minimal so that the functional 
  dependence of the predicted distribution on $y_3$ gives a good
  representation of the complete result.  This amounts to the
  principle of experimental optimization: fixing $\mu$ to the value $\mu
  = \mu_{\rm EO}$ which minimizes $\chi^2$ for that observable.
  Unless the differential two-jet rate is a lucky accident, the
  same principle should apply to any event-shape distribution (e.g.,
  thrust, energy-energy correlation, etc).  If using $\mu_{\rm EO}$
  does in fact subsume the effects of neglected terms, the values of
  $\als$ obtained from different event shape distributions, each at its own
  $\mu_{\rm EO}$, should be the same up to statistical and systematic
  errors other than those associated with scale uncertainty.  Burrows
  has analyzed various scale-fixing proposals\cite{burrows:warsaw} and
  finds that the {\it dispersion} of values of $\als$ obtained with the
  experimentally-optimized-scale prescription is no better than the {\it
  dispersion} from other scale-fixing prescriptions, as evident in
  Fig. \ref{burrows1} from ref. \cite{burrows:warsaw} reproduced here.
That study used $O(\alpha_s^2)$ rather than resummed predictions.
  However from Figs 34-37 of ref. \cite{sld:alphas} one can see that
  the dispersion in $\als$ determinations using $\mu_{EO}$ is large in
  this case as well, even restricting to matching schemes and
  observables for which $\chi^2/N_{dof} \le 1$.   Thus it is {\it not}
  true that the effects of neglected higher   order corrections are
  effectively included by choosing the $\mu$   which minimizes
  $\chi^2$.  

Next consider the problem of resummation scheme ambiguity.
  The log-R matching scheme gives the best-fit, with a $\chi^2$ of 78.5
  compared to $81.6$ from the R-matching scheme.  Therefore, if 
  the Bayesian procedure were being consistently implemented, the log-R
  matching scheme would be {\it a posteriori} identified as the
  model which gives the best estimate of the physical parameters.
  Within the set of models considered by ALEPH, their analysis
  procedure thus gives $n_f = 3.68 \pm 0.29 (stat) \pm 1.15(syst)$ as
  discussed above.  The $ \pm 0.29 $ statistical error means that
  repeating the experiment many times will yield a value of $n_f$ in
  the range 2.93-4.42 in 99\% of the trials.  The log-R   matching scheme
  prediction simply does not tolerate $n_f> 4.5$, for any 
  $\mu$, as apparent from Fig. \ref{aleph:mudep}.  Thus strictly
  following Bayesian reasoning\cite{dagostini:lec} requires reject the
  log R matching scheme for the $D_2$ observable because it
  contradicts our {\it a priori} knowledge that $n_f \ge 5$.  This
  phenomenon illustrates that Bayesian reasoning can only be applied
  safely when all 
  relevant experimental information is included in the fit and a
  sufficiently complete set of theoretical models is being fit.
  Otherwise, one can wind up in a local rather   than global minimum
  in the space of possible models.   

Hadronization errors are similarly not amenable to the Bayesian
method, unless much more general hadronization models are used.  In
particular, the parton shower modeling must include light gluinos and
the hadrons containing them, if one is to consistently determine
whether a model with light gluinos can fit the data.  In the absence of
such an undertaking, one can apply the traditional method of trying
various hadronization models and seeing if the resultant values of
$n_f$ change.  OPAL\cite{opal:4j95} investigated the effect on 4-jet
angular distributions of changing parameters in a given hadronization
model by $O(10\%)$ and found it produced shifts in $n_f$ of order 1
unit (see Table 1 of \cite{opal:4j95}).  ALEPH did not try changing
parameters in a given hadronization model, but reports that using Herwig
rather than JETSET to model the differential 2-jet rate leads to a
best-fit value of $n_f = 6.21$ with only a slight loss of fit quality
($\chi^2/N_{dof}$ increases from $81.6/73 = 1.12$ to $91.6/73 =
1.25$). The fact that such a large shift in $n_f$ accompanies a change
of model should have been followed up by varying $\mu$ and changing
matching scheme, to discover the best fit which can be obtained
with the new hadronization model.

\begin{figure}
\epsfxsize=\hsize
\epsffile{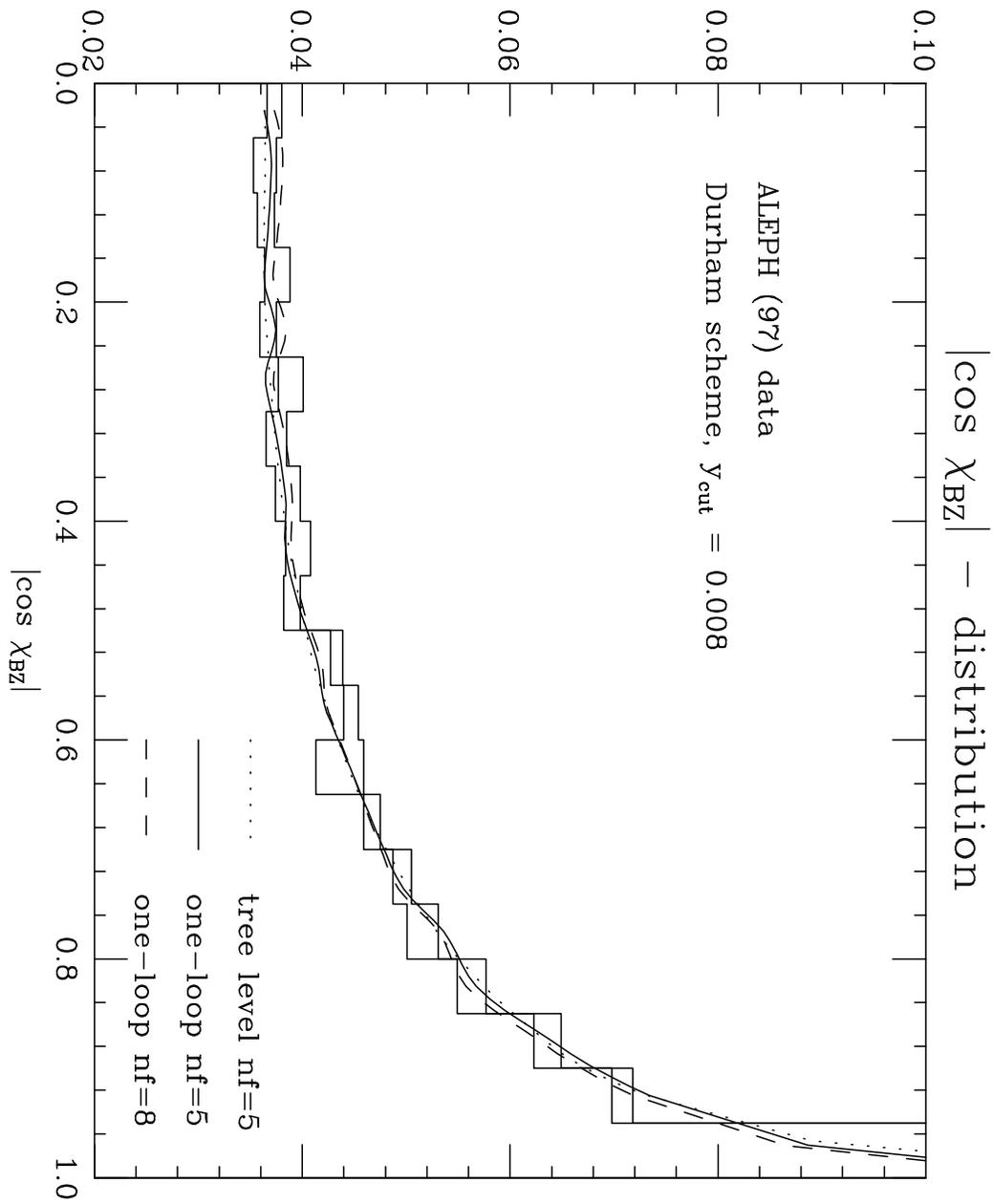}
\caption{Bengtsson-Zerwas angular distribution at tree and 1-loop level
  for $n_f = 5$ (dotted and solid) and at 1-loop for $n_f = 8$,
  compared to ALEPH data, from \cite{signer:moriond97}.}
\label{dixsigbz}
\end{figure}

A final area where there may be additional theoretical uncertainty
which cannot yet be quantified is the question of the sensitivity of
the 4-jet distributions to the scale $\mu$ and to details of the jet
definition ($y_{cut}$ and scheme).  The 1-loop correction to 4-jet
matrix elements was not available when the ALEPH analysis was started.
However it has recently been determined\cite{nlo} and the
correction to the angular dependence is small.  This can be seen in
Fig. \ref{dixsigbz} from ref. \cite{signer:moriond97} which shows the
Bengtsson-Zerwas angular distribution at tree level and 1-loop for
$n_f = 5$ and at 1-loop for $n_f = 8$, compared to ALEPH data.  Since
1-loop corrections to the angular distributions are small, $\mu$
dependence of the angular distribution is evidently not a problem.  At
the same time, Fig. \ref{dixsigbz} illustrates that light gluinos make
only a very small change in the angular distribution, which is why
the $n_f$ extraction is so sensitive to corrections to the
perturbative prediction. 

No resummed calculation exists for the 4-jet angular distributions,
but the effects of resummation may be large.  First of all, L3 found
considerable sensitivity to $y_{cut}$ in their color factor
analysis\cite{L3:ycut}.  Secondly, differential distributions such as
$D_2$ which depend on defining jets, for which the $y_{cut}$
dependence has been determined, display a significant sensitivity to
$y_{cut}$.  Finally, the sensitivity of $n_f$ to hadronization model
shows that low $p_t$ aspects of the distribution of hadrons relative
to the jets affect the angular distribution.  Thus soft and colinear
gluon radiation may modify the angular dependence found in a fixed
order calculation.    

It is desirable to estimate what result ALEPH would have reported had
traditional error estimation methods of earlier
experiments\cite{opal:4j95,delphi:tagged4j,L3:ycut} been used.  We can
only give a lower bound on the systematic error, since ALEPH did not
check the sensitivity of its $n_f$ extraction to variation of the
parameters of the hadronization model or choice of $y_{cut}$ for the
4-jet angular distributions, while the former made a significant
contribution to, e.g., the OPAL error\cite{opal:4j95}.  We must use R 
matching, since the log R matching scheme is excluded because it
contradicts the existence of the b quark.  We take $\sigma^{th} =
2.5$ because this reproduces the $n_f$ range obtained for $1/2 < \mu/M_Z
< 2$.  Combining quadratically with the $\pm 0.45 (had) \pm 0.27
(de) \pm 0.29 (stat)$ obtained by ALEPH, gives $n_f = 4.88 \pm 0.29
(stat) \pm 2.57 (syst)$.  Using the PDG prescription for obtaining
limits when part of the experimental range is unphysical ($n_f < 5$),
this implies the 95\% cl limit $n_f \le 9.95$, or the 77\% cl limit
$n_f \le 8$.  

It is interesting that the $\pm 2.57(syst)$ obtained from the ALEPH
combined fit to $D_2$ and the 4-jet angular distributions by
traditional uncertainty estimation procedures is actually a larger
than obtained using the 4-jet angular distributions
alone\cite{opal:4j95,delphi:tagged4j,L3:ycut}. The reason for this is that 
$D_2$ is actually very sensitive the $\mu$, even after resummation,
while the 4-jet angular distribution is not, so that the ALEPH
combined analysis only results in a more precise determination of
$n_f$ if the $\mu$ range can be limited.  However the results of
ref. \cite{burrows:warsaw} show that ALEPH's attempt to limit the
$\mu$ range by embracing the ``experimental optimization''
prescription is not valid.  

Csikor and Fodor (CF) receintly studied the effect of light gluinos on
the running of $\alpha_s(Q^2)$\cite{csikor_fodor}.  By restricting
themselves to $\alpha_s$ extracted via 3-loop perturbative predictions
for $R_{had}$ in $\tau$ decay and in $e^+ e^-$ collisions at the $Z^0$
and below, they largely avoid the hadronization and scale
uncertainties of the hadronic event shape analyses.  The fit is mainly
constrained by the values $\alpha_s(m_\tau) = 0.335 \pm 0.053$ and
$\als = 0.123 \pm 0.006$ because the data at intermediate values of
$Q$ are statistically much weaker.  However the error on
$\alpha_s(m_\tau)$ has been argued\cite{shifman96} to be of order
twice as large as used by CF.  Thus in their final analysis for the very
light gluino case, CF do not use $\alpha_s(m_\tau)$ at all.  They find
a very light gluino is disfavored only at about 70\% cl.  The 99.97\%
cl limit they quote after combining with the ALEPH limit mainly
reflects the ALEPH number, 
which we have argued underestimates the theoretical uncertainty and
does not correctly deal with the manifestly unphysical log R matching.
Modifying the ALEPH result to $n_f = 4.88 \pm 0.29 (stat) \pm 2.57
(syst)$, as obtained using standard methods of error estimation, leads
to the conclusion that very light gluinos cannot be excluded with
either the ALEPH or Csikor-Fodor methods until QCD uncertainties, which
are at this point mainly non-perturbative, can be reduced.

Before closing, I remark that SUSY model-building developments and
recent cosmological constraints have increased the importance of such
indirect constraints on light gluinos.  Several gauge-mediated
SUSY-breaking models have recently appeared in which the gluino is
lighter than the other gauginos and $R$-parity is
conserved\cite{gmsb}.  In such scenarios the only
decay mechanism for the lightest gluino-containing hadron (the
glueballino, denoted $R^0$) is through gravitino emission, if
kinematically allowed.  Thus the $R^0$ may be stable, or so long-lived
that direct detection through its decay may be impossible.
Furthermore, direct detection of decaying $R^0$'s may also be more
difficult than anticipated\cite{f:104} in models in which the photino
as well as the gluino is approximately massless at tree level.  In
this case, radiative corrections give the gluino and photino masses of
$\sim 0.1$ GeV and $\lsi 1$ GeV respectively, and the $R^0$ mass should be
1.3-2.2 GeV\cite{f:104}.  A new analysis\cite{f:113} of the photino
relic abundance shows that $m(\pho) \ge m(R^0)/1.8$.  Combining this with
the 1.3-2.2 GeV mass range of the $R^0$ leads to discouraging
prospects for direct searches for the decay $R^0 \rightarrow \pi^+
\pi^- \pho$ over the full $r \equiv m(R^0)/m(\pho)$ range in which
photinos provide the observed dark matter: $1.4 \lsi r \lsi
1.7$\cite{f:113}.  The point is that the invariant mass of the dipion
system cannot be larger than $m(R^0) - m(\pho) = m(R^0) (r-1)/r$,
while to remove the $K^0 \rightarrow \pi^+ \pi^-$ and other background
one wishes to require $m(\pi^+ \pi^-) $ convincingly above $m_K$.
Also, the low $Q$ value makes it more difficult to use the $p_t$ of
the $\pi^+ \pi^-$ pair as a discriminant against background. 

Thus even if there are light unstable gluino-containing hadrons, they
may be very difficult to observe through their decays.  This increases
the importance of reducing the (mainly non-perturbative) QCD errors in
theoretical predictions for $\alpha_s$ running and for $e^+ e^-$ event
shape observables.  When lattice gauge theory can predict the
bottomonium spectrum without the use of non-relativistic or quenched 
approximations, the $n_f$ required to obtain precision agreement
for the level splitting may provide the cleanest indirect test of
all.  Direct searches for quasi-stable R-hadrons as discussed in
\cite{f:51,f:95,f:104} should be considered.

My La Thuile talk ended with a brief discussion of three tantalizing
phenomena which could be naturally explained by light gluinos but are
seemingly difficult to explain otherwise.  Here, I merely list them in
order of decreasing experimental robustness and leave the reader to
consult the references given for details and further references.
\begin{itemize}
\item  The existence of the isosinglet pseudoscalar
  $\eta(1410)$ whose properties match those expected for a
  gluinoball or glueball and whose mass agrees with that predicted for
  a gluinoball but is much lighter than the lattice QCD and sum rule
  predictions for a glueball\cite{f:109}.
\item  Ultra-high-energy cosmic rays which violate the
  Greisen-Zatsepin-Kuzmin bound\cite{f:115,f:114}.
\item  The peak observed in dijet mass pairs by ALEPH in $e^+ e^-$
  annihilation at 133, 161 and 172 GeV.
\end{itemize}

{\bf Acknowledgements:} 
I am particularly indebted to G. Dissertori for extensive correspondence
and discussions of the ALEPH analysis.  I have also benefited from
correspondence and discussions with S. Bentvelsen, P. Burrows,
S. Catani, G. Cowan, L. Dixon, Z. Fodor, J. W. Gary, M. Seymour and
D. Schlatter.  Special thanks to Dissertori, Burrows, Dixon and Signer
for providing their figures.




\end{document}